\documentclass[11pt]{article}
\usepackage[utf8]{inputenc}
\usepackage[UKenglish]{babel}
\usepackage{a4wide}
\usepackage[colorlinks=false]{hyperref}
\usepackage{amsfonts}
\usepackage{amssymb}
\usepackage{graphics}
\usepackage{amsmath}
\usepackage{cite}
\usepackage{appendix}
\usepackage{etoolbox}
\usepackage{physics}
\usepackage{sectsty}

\DeclareMathOperator{\correspondencia}{AdS_{5}/CFT_{4}}
\DeclareMathOperator{\variedaduno}{AdS_{5}\times S^{5}}

\DeclareMathOperator{\variedaddos}{AdS_{3}\times S^{3}\times T^{4}}
\DeclareMathOperator{\variedadtres}{AdS_{3}\times S^{3}}
\DeclareMathOperator{\variedadcuatro}{AdS_{3}\times S^{1}}
\DeclareMathOperator{\variedadcinco}{\mathbb{R}\times S^{3}}
\DeclareMathOperator{\adstres}{AdS_{3}}
\DeclareMathOperator{\stres}{S^{3}}
\DeclareMathOperator{\sdos}{S^{2}}
\DeclareMathOperator{\suno}{S^{1}}
\DeclareMathOperator{\tcuatro}{T^{4}}
\DeclareMathOperator{\hdos}{H_{2}}
\DeclareMathOperator{\U}{U(1)}
\DeclareMathOperator{\SU}{SU(2)}
\DeclareMathOperator{\PSL}{PSL(2,\mathbb{R})}
\DeclareMathOperator{\SL}{SL(2,\mathbb{R})}
\DeclareMathOperator{\SLL}{SL(2,\mathbb{R})_{L}}
\DeclareMathOperator{\SLR}{SL(2,\mathbb{R})_{R}}

\DeclareMathOperator{\SLLRdos}{SL(2,\mathbb{R})_{L}\times SL(2,\mathbb{R})_{R}/SL(2,\mathbb{R})}
\DeclareMathOperator{\SO}{SO(6)}
\DeclareMathOperator{\SUeszett}{SU(1|1)}
\DeclareMathOperator{\PSU}{PSU(1,1|2)}
\DeclareMathOperator{\dif}{d\!}
\DeclareMathOperator{\e}{e}
\DeclareMathOperator{\im}{i}

\DeclareMathOperator{\ie}{i.\! e.}
\DeclareMathOperator{\viz}{viz.}
\DeclareMathOperator{\cf}{cf.}

\def\texto
       {\topmargin -20pt   
        \oddsidemargin 0pt
        \headheight 0pt 
        \headsep 0pt
        \textwidth 6.35in       
        \textheight 9.25in      
        }
\texto

\begin{document}

\begin{titlepage}

\begin{center}

\vskip 0.25in

{\LARGE The~$\SL$ Wess-Zumino-Novikov-Witten spin-chain $\sigma$-model}

\vskip 0.25in
     
{\bf Roberto Ruiz} 
\vskip 0.1in

Departamento de F{\'i}sica Te{\'o}rica \\
and \\
Instituto de F{\'i}sica de Part{\'i}culas y del Cosmos (IPARCOS), \\
Universidad Complutense de Madrid, \\
$28040$ Madrid, Spain \\
{\footnotesize{\texttt{roruiz@ucm.es}}}

\end{center}

\vskip .4in

\centerline{\bf Abstract}

\vskip .1in

\noindent The~$\SL$ Wess-Zumino-Novikov-Witten model realises bosonic-string theory in~$\adstres$ with pure Neveu-Schwarz-Neveu-Schwarz flux. We construct an effective action in the semi-classical limit of the model, which corresponds to a~$\SL$ spin-chain $\sigma$-model. We adopt two complementary points of view. First, we consider the classical action. We identify fast and slow target-space coordinates. We impose a gauge-fixing condition to the former. By expanding the gauge-fixed action in an effective coupling, we obtain the effective action for the slow coordinates. Second, we consider the spin chain of the model. We postulate a set of coherent states to express a transition amplitude in the spin chain as a path integral. We observe that the temporal interval is discretised in terms of the step length of the spatial interval. This relationship implies that the Landau-Lifshitz limit of the spin chain involves both intervals. The limit yields a semi-classical path integral over coherent states, wherein we identify the effective action again.

\end{titlepage}

\vfill
\eject

\def\baselinestretch{1.2}
\baselineskip 15pt
\sectionfont{\large}
\renewcommand{\theequation}{\thesection.\arabic{equation}}
\csname @addtoreset\endcsname{equation}{section}

\section{Introduction}

\label{seccionuno}

\noindent The~$\SL$ Wess-Zumino-Novikov-Witten~(WZNW) model provides a testing ground for non-perturbative techniques due to its exact solvability.~\footnote{We denote by~$\SL$ and~$\PSU$ the universal cover of their Lie group and supergroup, respectively.} The model realises bosonic-string theory in~$\adstres$ supported with pure Neveu-Schwarz-Neveu-Schwarz~(NSNS) three-form flux. The spectrum of the~$\SL$ WZNW model is computable by analysing the representations of the  Ka{\v{c}}-Moody algebra of the currents in the world-sheet two-dimensional conformal field theory~(CFT$_{2}$)~\cite{0001053}. The~$\PSU$ WZNW model is the supersymmetric embedding of the model, and realises type IIB superstring theory on~$\variedadtres\subset\variedaddos$ with pure NSNS flux~\cite{9902098}. In order for the~$\PSU$ WZNW model to account for the supersymmetric embedding in~$\tcuatro\subset\variedaddos$, the world-sheet CFT$_{2}$ includes supplementary bosonic and fermionic field operators; see, for example,~Section 2 of \cite{1704.08667}. The spectrum of the~$\PSU$ WZNW model is computable through the study of representations of the current algebra~\cite{0610070}. 

Non-linear $\sigma$-models on highly symmetric backgrounds are often quantum integrable, and the~$\SL$ WZNW model is no exception. It was both anticipated in~\cite{1804.01998} and established in~\cite{1806.00422} that the spectrum of the~$\PSU$ WZNW model is retrievable from an integrable spin chain. The spin chain encodes the spectrum in a set of Bethe equations. The cancellation of wrapping corrections in the thermodynamic Bethe ansatz renders the Bethe equations exact. Furthermore, the explicit resolution of the equations turns out to be feasible. The Bethe equations thus supply the spectrum in representations of the current algebra built upon the principal discrete series of~$\PSU$, and hence~$\SL$. In addition, the Bethe equations admit exceptional solutions which violate the unitarity bound for representations of the principal discrete series. Exceptional solutions were argued to reproduce the spectrum connected to the principal continuous series of~$\PSU$ and~$\SL$ in~\cite{2010.02782}.

The integrable system advanced in~\cite{1804.01998,1806.00422} poses the question of the emergence of the spin chain in the semi-classical limit of the~$\PSU$ WZNW model. This question was answered in~\cite{1311.1794,1905.05533} on the basis of the proposal of~\cite{0311203}. Reference~\cite{0311203} put forward an effective action in the~$\SU$ sector of the~$\correspondencia$ correspondence, which is closed at all-loop order in the 't Hooft coupling. The effective action corresponds to a~$\SU$ spin-chain $\sigma$-model, and is retrievable from both members of the~$\correspondencia$ correspondence in the semi-classical limit. At leading order in the effective coupling, the agreement implies the matching of the spectra (and the hierarchies of conserved charges) in the~$\SU$ sector. The agreement also clarifies the appearance of string configurations in the spin chain by means of coherent states. In the~$\SU$ sector of type IIB superstring theory on the~$\variedaduno$ background with pure Ramond-Ramond flux, the derivation of~\cite{0311203} starts from the Polyakov action on~$\variedadcinco$. The procedure of~\cite{0311203} can be summarised in three steps~\cite{0403120}: the identification of fast and slow coordinates, the imposition of a gauge-fixing condition on fast coordinates, and the expansion of the Polyakov action in an effective coupling in the semi-classical limit. The expansion is permitted by the generalised velocities of the slow coordinates since they are suppressed by the effective coupling. The result is an effective action for the slow coordinates. This action is retrievable from the dual~$\mathcal{N}=4$ supersymmetric Yang-Mills theory on the conformal boundary of the~$\variedaduno$ background. The derivation of \cite{0311203} begins with the XXX$_{1/2}$ Heisenberg model in this case. This spin chain encodes the spectrum in the~$\SU$ sector of the dual theory at one-loop order in the 't Hooft coupling. The effective action is obtained in two stages. First, a general transition amplitude is expressed as an exact path integral over coherent states. Then, the Landau-Lifshitz (LL) limit is applied to the action within the path integral. The LL limit is a continuum limit on the spatial interval of the spin chain; by construction, it is semi-classical. The LL limit produces an effective action for the coordinates that parameterise coherent states and matches the aforementioned effective action under the identification of coherent-state coordinates with slow coordinates. The effective action is linear in the generalised velocities, but it is quadratic in the spatial derivatives. For a review of spin-chain $\sigma$-models in the $\correspondencia$ correspondence, the reader is referred to~\cite{0409296}.

References~\cite{1311.1794,1905.05533} modified the method of~\cite{0311203} and applied it to the~$\PSU$ WZNW model. Specifically,~\cite{1311.1794,1905.05533} constructed an effective action in the truncation of the model to the~$\SU$ WZNW model, which realises bosonic-string theory in~$\stres$ with pure NSNS flux. The effective action corresponds to a~$\SU$ spin-chain $\sigma$-model that is linear in both the generalised velocities and the spatial derivatives. The derivation of~\cite{1311.1794} starts from the classical action of the~$\SU$ WZNW model, $\ie$ the Polyakov action on~$\variedadcinco$ supplied with a Wess-Zumino (WZ) term for the B-field in~$\stres$. (The derivation instead starts from a non-linear $\sigma$-model on the integrable background of~\cite{1209.4049}, which is a general deformation of the background considered here.) The derivation first distinguishes between fast and slow target-space coordinates. The former set is determined in~\cite{1311.1794} by the equations of motion and the Virasoro constraints. The generalised velocities of the slow coordinates are not suppressed by the effective coupling in the semi-classical limit; the B-field implies that the generalised velocities appear at the same order as the spatial derivatives. The expansion in the effective coupling of the gauge-fixed action supplies an effective action for the slow coordinates. The starting point of~\cite{1905.05533} is the~$\SU$ sector of the spin chain of~\cite{1804.01998,1806.00422}. The derivation of~\cite{1905.05533} contrasts with~\cite{0311203}. First, the temporal interval is discretised in terms of the spatial step length. Second, the LL limit is not a continuum limit in the spatial interval applied to the action of an exact path integral. The LL limit is a simultaneous continuum limit on the temporal and spatial intervals. It brings forth a representation of the transition amplitude as a semi-classical path integral over coherent states. Coherent states are in fact not known a priori in \cite{1905.05533}, but are postulated beforehand. In this case, the action within the path integral is the effective action. If coherent states are properly parameterised, the effective actions of~\cite{1311.1794} and~\cite{1905.05533} match. 

The semi-classical agreement in the~$\SU$ WZNW model raises the question of the availability of a spin-chain $\sigma$-model in other truncations of the~$\PSU$ WZNW model. In this article, we construct a~$\SL$ spin-chain $\sigma$-model in the semi-classical limit of the~$\SL$ WZNW model. The~$\SL$ spin-chain $\sigma$-model in the~AdS$_{5}$/CFT$_{4}$ correspondence was analysed in~\cite{0404133,0409086,0501203}, some of whose elements we borrow. (Solutions to the equations of motion of the model were presented in~\cite{0409086,0409217,0410226}.) The article is composed of the following sections. In Section~\ref{secciondos}, we derive an effective action from the classical action of the~$\SL$ WZNW model. We divide target-space coordinates into fast and slow coordinates by calling on the light-cone gauge-fixing condition of~\cite{1806.00422}. We impose the static gauge-fixing condition to the fast coordinates by means of the formal T-duality of~\cite{0406189,0501203}. This condition reveals that, in the semi-classical limit, the generalised velocities and the spatial derivatives of the slow coordinates appear at the same order in the effective coupling. We rephrase the gauge-fixed action as a Nambu-Goto (NG) form with a WZ term, allowing us to obtain an effective action that is linear in both the generalised velocities and the spatial derivatives through an expansion in the effective coupling. We then study some solutions previously considered in the bibliography. In Section~\ref{secciontres}, we derive the effective action starting from the~$\SL$ sector of the spin chain of~\cite{1804.01998,1806.00422}. First, we present the spin chain, then we postulate a set of coherent states. Except for a slight generalisation, we mimic~\cite{1905.05533} to rephrase the transition amplitude in the spin chain as a semi-classical path integral over the coherent states. The temporal interval is discretised in terms of the spatial step length. The LL limit is a synchronised continuum limit on the temporal and spatial intervals, and provides the path integral in the semi-classical limit. Some manipulations allow us to recover the effective action of Section~\ref{secciondos}. In Section~\ref{seccioncuatro}, we close with a summary and comment on open problems.

\section{The \texorpdfstring{$\mathrm{SL}(2,\mathbb{R})$}{SL(2,R)} spin-chain \texorpdfstring{$\sigma$}{sigma}-model from the classical action}

\label{secciondos}

In this section, we derive the action of the~$\SL$ spin-chain $\sigma$-model from the classical action of the~$\SL$ WZNW model coupled to a scalar field parameterising the equator~$\suno$ of~$\stres$. 
Our methodology builds upon~\cite{0406189}, which is a refinement of~\cite{0311203}. Reference~\cite{0406189} studied the~$\SU$ sector of the~$\variedaduno$ background. Reference~\cite{0501203} applied the methodology of~\cite{0406189} to the~$\SL$ sector thereof.
 
To begin, we must divide the target-space coordinates into two sets: `fast' and `slow' coordinates. Fast coordinates consist of a time-like coordinate and a space-like coordinate whose generalised velocities are large in the semi-classical limit~$k\rightarrow\infty$. Fast coordinates can be determined through the static gauge-fixing condition under the formal T-dualisation of the space-like coordinate. In this way, fast coordinates parameterise the temporal and spatial directions in the continuum limit of the spin chain. Slow coordinates are the remaining coordinates. In the~$\variedaduno$ background, the generalised velocities of slow coordinates are suppressed when~$k\rightarrow\infty$. The presence of the B-field modifies this property by means of the spatial derivatives of the slow coordinates. Slow coordinates parameterise coherent states in the continuum limit of the spin chain.

Therefore, we must introduce a parameterisation in~$\variedadcuatro$ to get a split into fast and slow coordinates. The parameterisation of~$\adstres$ is provided by the Hopf fibration~\cite{0404133,0501203}. The fibration reveals the local resemblance between~$\hdos\times\suno$ and~$\PSL$, of which~$\adstres$ is the universal cover. The resemblance is a consequence of the fibre-bundle structure of~$\PSL$, whereby~$\hdos$ is the base space and~$\suno$ is the fibre. The gauge group of the fibre bundle is~$\U$. The Hopf fibration is conveniently expressed in the embedding coordinates of~$\adstres$ into~$\mathbb{R}^{2,2}$. Let us denote them by~$Y^{A}$, with~$A=0,1,2,3$. The embedding coordinates satisfy~$\eta_{AB}Y^{A}Y^{B}=-1$, where~$\eta_{AB}=\textnormal{diag}(-,+,+,-)$. The Hopf fibration corresponds to the parameterisation
\begin{equation}
\label{ecuaciondosuno}
Y^{0}+\im Y^{3}=\e^{\im\alpha}Z_{0} \ , \quad \ Y^{1}+\im Y^{2}=\e^{\im\alpha}Z_{1} \ .
\end{equation}
The coordinate~$\alpha$ is real and parameterises the fibre $\suno$, and~$Z_{a}$ are complex coordinates subject to the constraint~$-\overline{Z_{0}}Z_{0}+\overline{Z_{1}}Z_{1}=-1$. The action of $\U$ on $\alpha$ and $Z_{a}$ reads~$\alpha\mapsto\alpha+\beta$ and~$Z_{a}\mapsto\exp(-\im\beta)Z_{a}$, hence preserves the constraint of $Z_{a}$. Therefore,~$Z_{a}$ parameterise the base space~$\hdos$. By construction, the coordinate~$\alpha$ is fast, whereas~$Z_{a}$ are~slow. 

In Section~\ref{secciontres}, the effective action that arises in the semi-classical of the spin chain is not expressed in~(\ref{ecuaciondosuno}). Instead, it is written in the global coordinate system of~$\adstres$, which is related to $Y^{A}$ as
\begin{equation}
\label{ecuaciondosdos}
Y^{0}+\im Y^{3}=\e^{\im t}\cosh\rho\ , \quad \ Y^{1}+\im Y^{2}=\e^{\im\psi}\sinh\rho \ ,
\end{equation}
where~$t\in(-\infty,\infty)$,~$\rho\in[0,\infty)$ and~$\psi\in[0,2\pi)$. Notice that, being non-compact, the range of~$t$ forbids closed time-like curves in~$\adstres$. The comparison with Section~\ref{secciontres} requires us to identify~$\alpha$ and~$Z_{a}$ in~(\ref{ecuaciondosdos}). However, the relationship between~(\ref{ecuaciondosuno}) and~(\ref{ecuaciondosdos}) is not straightforward due to the action of~$\U$ on $\alpha$ and $Z_{a}$.

To resolve the ambiguity, we call on the the gauge-fixing condition for the fast coordinates with respect to the world-sheet coordinates. The procedure of~\cite{0406189} involves the imposition of the static gauge-fixing condition to the fast coordinates. By definition, this condition fixes~$\alpha$ proportionally to the time-like world-sheet coordinate~$\tau$. Each admissible identification of~$\alpha$ in~(\ref{ecuaciondosdos}) leads to a gauge-fixing condition. However, we need to choose~$\alpha$ properly for the comparison with the semi-classical limit of the spin chain. Reference \cite{1806.00422} built the spin chain on the basis of the light-cone gauge-fixing condition, where~$t$ is proportional to~$\tau$. We must then ensure that the static and light-cone gauge-fixing conditions are compatible. In this way, we determine
\begin{equation}
\label{ecuaciondostres}
\alpha=t \ , \quad Z_{0}=\cosh\rho \ , \quad Z_{1}=\e^{\im\varphi}\sinh\rho \ ,
\end{equation}
where~$\varphi=\psi-t$. The coordinate~$t$ is thus the time-like fast coordinate, and~$\rho$ and~$\varphi$ are slow coordinates. It is worth noting that these identifications were introduced in the~$\variedaduno$ background to construct the~$\SL$ spin-chain $\sigma$-model~\cite{0409086,0404133,0501203}.

In addition, we must identify the space-like fast coordinate. In general, the Hopf fibration singles this coordinate out in~$\stres$~\cite{0403120,0406189}. The Hopf fibration manifests the fibre-bundle structure of~$\stres$, whereby~$\sdos$ is the base space and~$\suno$ is the fibre. In~$\variedadcuatro$, the base space of~$\stres$ is collapsed to a point. Therefore, the space-like fast coordinate is the coordinate~$\phi$ of the equator~$\suno\subset\stres$.

Once we have identified target-space coordinates properly, we can start the derivation of the effective action. The starting point is the Polyakov action on~$\variedadcuatro$ with a WZ term for the B-field in $\adstres$. The associated Lagrangian is
\begin{equation}
\label{ecuaciondoscuatro}
\begin{split}
L=&-\frac{k}{4\pi}\left[\gamma^{\alpha\beta}(-\partial_{\alpha}t\partial_{\beta}t+\partial_{\alpha}\rho\partial_{\beta}\rho+2\sinh^{2}\rho\,\partial_{\alpha}t\partial_{\beta}\varphi +\sinh^{2}\rho\,\partial_{\alpha}\varphi\partial_{\beta}\varphi +\partial_{\alpha}\phi\partial_{\beta}\phi\right.) \\
&-2\epsilon^{\alpha\beta}\sinh^{2}\rho\,\partial_{\alpha}t\partial_{\beta}\varphi\left.\vphantom{\gamma^{\alpha\beta}}\right] \ .
\end{split}
\end{equation}
Above,~$k\in\mathbb{N}$ denotes the level,~$\gamma_{\alpha\beta}$ denotes a unimodular world-sheet metric whose signature is~$(-,+)$,~$\epsilon^{\alpha\beta}$ denotes the Levi-Civita symbol with~$\epsilon^{\tau\sigma}=-\epsilon^{\sigma\tau}=-1$, and lower-case Greek indices run over~$\tau\in(-\infty,\infty)$ and the space-like world-sheet coordinate~$\sigma\in[0,2\pi)$. We emphasise that no gauge-fixing condition has been imposed on~$\gamma_{\alpha\beta}$. The fields in~(\ref{ecuaciondoscuatro}) satisfy closed-string boundary conditions:
\begin{equation}
\label{ecuaciondoscinco}
\begin{split}
t(\tau,\sigma+2\pi)&=t(\tau,\sigma) \ , \quad \rho(\tau,\sigma+2\pi)=\rho(\tau,\sigma) \ , \\
\varphi(\tau,\sigma+2\pi)&=\varphi(\tau,\sigma)+2\pi m \ , \quad \phi(\tau,\sigma+2\pi)=\phi(\tau,\sigma)+2\pi n \ .
\end{split}
\end{equation}
The indices~$m,n\in\mathbb{Z}$ are winding numbers. In particular,~$n$ determines the level-matching condition~\cite{1806.00422}. 

At this stage, we must impose a gauge-fixing condition to the fast coordinates. Let us focus on~$\phi$ first. The light-cone gauge-fixing condition for~$\phi$ in~\cite{1806.00422} is~$p_{\phi}=J/2\pi$, where~$p_{\phi}$ denotes the canonically conjugate momentum of~$\phi$ and~$J$ denotes the total angular momentum:
\begin{equation}
\label{ecuaciondosseis}
J=-\frac{k}{2\pi}\int_{0}^{2\pi}\dif\sigma\gamma^{\tau\alpha}\partial_{\alpha}\phi \ .
\end{equation}
As we have mentioned before, we look for a static gauge-fixing condition relative to~$\phi$ that is equivalent to~$p_{\phi}=J/2\pi$. One may be tempted to fix~$\phi$ proportionally to~$\sigma$, but this relationship is not consistent with~(\ref{ecuaciondosseis}). Moreover, the direct usage of~$p_{\phi}=J/2\pi$ would require the Hamiltonian formalism~\cite{0403120,0406189}. We can both find a static gauge-fixing condition equivalent to~$p_{\phi}=J/2\pi$ and avoid the Hamiltonian formalism by resorting to the formal T-dualisation of~$\phi$ into~$\widehat{\phi}$~\cite{0406189,0501203}. 

We T-dualise~$\phi$ into~$\widehat{\phi}$ on the basis of the Buscher procedure; see, for example,~\cite{9410237}. Accordingly, we replace~$\dif\phi$ by a one-form~$A$ in~(\ref{ecuaciondoscuatro}), and add a term to such the Lagrangian:
\begin{equation}
\label{ecuaciondossiete}
L\mapsto L-\frac{k}{2\pi}\widehat{\phi}\,\epsilon^{\alpha\beta}\partial_{\alpha}A_{\beta} \ .
\end{equation}
The Lagrange multiplier~$\widehat{\phi}$ ensures that~$A$ is closed, $\ie$~$\dif A=0$. Being closed,~$A$ is locally exact, which means that~$A=\dif\phi$ locally. If exactness held globally, we could integrate~$\widehat{\phi}$ out in~(\ref{ecuaciondossiete}) and~$A=\dif\phi$ would provide~(\ref{ecuaciondoscuatro}) again. Nonetheless, the world-sheet is cylindrical, its first Betti number equals one, and~$A$ may not be globally exact. We overcome this obstruction by arguing that T-duality is formal in this case. Formal T-duality allows us to derive the correct action of the~$\SL$ spin-chain~$\sigma$-model without need for the Hamiltonian formalism. We thus ignore topological issues concerning the global exactness of~$A$. We also pass over the dual quantisation of~(\ref{ecuaciondosseis}) and the dilatonic contribution from the path-integral measure (which would vanish in any case).

Bearing in mind the previous caveat, we eliminate $A$ from~(\ref{ecuaciondossiete}) by solving its equations of motion. We obtain
\begin{equation}
\label{ecuaciondosocho}
A_{\alpha}=-\epsilon_{\alpha\beta}\gamma^{\beta\delta}\partial_{\delta}\widehat{\phi} \ .
\end{equation}
If we implement this expression in~(\ref{ecuaciondossiete}), we arrive to the Lagrangian~(\ref{ecuaciondoscuatro}) up to the replacement of~$\phi$ by~$\widehat{\phi}$ (once we omit the contribution of a total derivative). The central improvement brought forth by formal T-duality is the reformulation of the expression of~$J$. The substitution of~(\ref{ecuaciondosocho}) in~(\ref{ecuaciondosseis}) leads us to
\begin{equation}
\label{ecuaciondosnueve}
J=\frac{k}{2\pi}(\widehat{\phi}(\tau,2\pi)-\widehat{\phi}(\tau,0)) \ .
\end{equation}
Therefore,~$p_{\phi}=J/2\pi$ renders into the static gauge-fixing condition~$\widehat{\phi}=(J/k)\sigma$. The coordinate~$\widehat{\phi}$ thus parameterises the spatial direction of the spin chain in the continuum limit. 

The semi-classical limit~$k\rightarrow\infty$ raises the question of the legitimacy of~$\widehat{\phi}=(J/k)\sigma$ as a regular gauge-fixing condition. The answer relies on the fact that~$\phi$ is fast by assumption. By definition, its generalised velocity~$v^{\phi}=\gamma^{\tau\alpha}\partial_{\alpha}\phi$ is large. Therefore, the light-cone gauge-fixing condition~$p_{\phi}=-(k/2\pi)v^{\phi}=J/2\pi$ implies that~$J$ is semi-classical and large, hence the ratio~$J/k$ is non-vanishing and large. Not only is~$k/J$ small, but~$k/J$ is also a sensible effective coupling in the semi-classical limit. 

In addition, we must impose the static gauge-fixing condition to~$t$. As we have discussed before (\ref{ecuaciondostres}), the light-cone gauge-fixing condition of~\cite{1806.00422} requires us to fix~$t=a\tau$. This choice reflects the coincidence of the time-like directions on the target-space and the world-sheet, as  well as the temporal direction of the spin chain in the continuum limit. Note that no obstruction forbids~$t=a\tau$ as the fields in the Lagrangian do not satisfy any explicit boundary conditions with respect to~$\tau$. We determine the proportionality coefficient~$a$ by arguing the observation presented in~\cite{0311203}, where it was emphasised that~$t$ is fast if~$a$ is large when~$k\rightarrow\infty$. (Consult~\cite{0409086} for the analogous discussion in the~$\SL$ sector of the~$\variedaduno$ background.) If we further insist on the Berestein-Maldacena-Nastase (BMN) scaling of the semi-classical space-time energy, $\ie$~that the leading contribution to the space-time energy is~$J$, we obtain~$a=J/k$.~\footnote{Alternatively, we could have followed~\cite{0406189,0501203}. One thus fixes~$t=\tau$, and rescales~$\tau$ afterwards in a NG action with a WZ term akin to~(\ref{ecuaciondosquince}).}

In sum, the static gauge-fixing condition is
\begin{equation}
\label{ecuaciondosdiez}
t=\frac{J}{k}\tau \ , \quad \widehat{\phi}=\frac{J}{k}\sigma \ .
\end{equation}
We can use the condition above in the equations of motion following from~(\ref{ecuaciondoscuatro}) (once~$\phi$ is replaced by~$\widehat{\phi}$). This step is convenient as the Euler-Lagrange equations supplies us with a cross-check to validate the equations of motion from the effective action. If we use~(\ref{ecuaciondosdiez}), the equations of motion of~(\ref{ecuaciondoscuatro}) read
\begin{align}
\label{ecuaciondosonce}
&\frac{J}{k}\partial_{\alpha}\gamma^{\alpha\tau}-\partial_{\alpha}[\,(\gamma^{\alpha\beta}-\epsilon^{\alpha\beta})\sinh^2\rho\,\partial_{\beta}\varphi\,]=0 \ , \\
\label{ecuaciondosdoce}
&\frac{J}{k}\sinh(2\rho)\,(v^\varphi-\varphi')-\partial_{\alpha}(\gamma^{\alpha\beta}\partial_{\beta}\rho)+\frac{1}{2}\sinh(2\rho)\gamma^{\alpha\beta}\partial_{\alpha}\varphi\partial_{\beta}\varphi=0 \ , \\
\label{ecuaciondostrece}
&\frac{J}{k}[\,\sinh^2\rho\,\partial_{\alpha}\gamma^{\alpha\tau}+\sinh(2\rho)\,(v^\rho-\rho')\,]+\partial_{\alpha}(\gamma^{\alpha\beta}\sinh^2\rho\,\partial_{\beta}\varphi)=0 \ , \\
\label{ecuaciondoscatorce}
&\frac{J}{k}\partial_{\alpha}\gamma^{\alpha\sigma}=0 \ , 
\end{align}
where~$v^{\rho}=\gamma^{\tau\alpha}\partial_{\alpha}\rho$ and~$v^{\varphi}=\gamma^{\tau\alpha}\partial_{\alpha}\varphi$ are the generalised velocities of~$\rho$ and~$\varphi$, respectively. The equations of motion are split in terms that are multiplied by~$J/k$ and terms that are not. In the semi-classical limit, both sets of equations must be satisfied separately because~$k/J$ is the effective coupling. At leading order in~$k/J$, equations~(\ref{ecuaciondosonce}) and~(\ref{ecuaciondoscatorce}) imply that~$\gamma^{\alpha\beta}$ is divergenceless. Since the unimodular world-sheet metric~$\gamma_{\alpha\beta}$ is covariantly constant with respect to the torsionless connection, we deduce~$\gamma^{\alpha\beta}=\eta^{\alpha\beta}+\textnormal{O}(k/J)$. This result simplifies the equations of motion of the slow coordinates~(\ref{ecuaciondosdoce}) and~(\ref{ecuaciondostrece}). Explicitly, it implies that~$\dot{\rho}+\rho'=\textnormal{O}(k/J)$ and~$\dot{\varphi}+\varphi'=\textnormal{O}(k/J)$ (as long as~$\rho$ does not vanish). The appearance of~$\rho'$ and~$\varphi'$ may be traced back to the presence of a non-trivial B-field in~$\variedadcuatro$. It is worth noting at this point that the generalised velocities of the slow coordinates in the~$\variedaduno$ background are not balanced by spatial derivatives, but they are suppressed by the effective coupling~\cite{0404133,0409086,0501203}. 
 
The next step we need to take is the reduction of the gauge-fixed action to a NG form with a WZ term. Let us thus rephrase the world-sheet metric as~$\gamma^{\alpha\beta}=\sqrt{-h}h^{\alpha\beta}$, where~$h_{\alpha\beta}$ is a general non-unimodular world-sheet metric and~$h$ is its determinant. We solve the Virasoro constraints through the identification of~$h_{\alpha\beta}$ with the induced metric on the world-sheet and obtain
\begin{equation}
\label{ecuaciondosquince}
S=-\frac{k}{2\pi}\int_{-\infty}^{\infty}\dif\tau\int_{0}^{2\pi}\dif\sigma\left(\sqrt{-h}-\frac{J}{k}\sinh^2\rho\,\varphi'\right) \ .
\end{equation}
The expression of the determinant~$h$ is arranged in powers of the inverse effective coupling $J/k$; explicitly,
\begin{equation}
\label{ecuaciondosdieciseis}
\begin{split}
h&=-\frac{J^{4}}{k^{4}}+2\frac{J^{3}}{k^{3}}\sinh^2\rho\,\dot{\varphi}-\frac{J^{2}}{k^{2}}\,[-\dot{\rho}^2+\rho'^2+\sinh^2\rho\,(-\dot{\varphi}^2+\cosh^2\rho\,\varphi'^2)\,] \\
&-2\frac{J}{k}\sinh^2\rho\,\rho'(\dot{\rho}\,\varphi'-\rho'\dot{\varphi})+\sinh^2\rho\,(\dot{\rho}\,\varphi'-\rho'\dot{\varphi})^2 \ .
\end{split}
\end{equation}
It follows that, just as the combination~$\lambda/J^2$ in the~$\variedaduno$ background~\cite{0204226,0304255}, the ratio~$k/J$ is the effective coupling that permits an analytical expansion.~\footnote{It may seem that~$k/J$ is not analytic as~$k$ is not quantised in the~$\SL$ WZNW model~\cite{0001053}. However, this is not the case:~$k$ is indeed quantised in the~$\PSU$ WZNW model that embeds the model~\cite{9902098}.} It is worth mentioning that~$\sqrt{-h}h^{\alpha\beta}=\eta^{\alpha\beta}+\textnormal{O}(k/J)$, which is consistent with~(\ref{ecuaciondosonce}) and~(\ref{ecuaciondoscatorce}).

Finally, let us expand~(\ref{ecuaciondosquince}) with respect to~$k/J$. If we neglect~$\textnormal{O}(k/J)$ terms and omit the divergent constant contribution of the fast coordinates, we obtain the following effective action:
\begin{equation}
\label{ecuaciondosdiecisiete}
S=\frac{J}{2\pi}\int_{-\infty}^{\infty}\dif\tau\int_{0}^{2\pi}\dif\sigma\sinh^2\rho\,(\dot{\varphi}+\varphi')\ .
\end{equation}
The effective action corresponds to a~$\SL$ spin-chain $\sigma$-model, and is linear in both~$\dot{\varphi}$ and~$\varphi'$. It is in exact agreement with the expression for the effective action~(\ref{ecuaciontresdiecinueve}) that is shown in Section~\ref{secciontres}. The equations of motion of~(\ref{ecuaciondosdiecisiete}) are
\begin{equation}
\label{ecuaciondosdieciocho}
\sinh(2\rho)(\dot{\rho}+\rho')=0 \ , \quad \sinh(2\rho)(\dot{\varphi}+\varphi')=0 \ ,
\end{equation}
which are consistent with~(\ref{ecuaciondosdoce}) and~(\ref{ecuaciondostrece}). The general solution to~(\ref{ecuaciondosdieciocho}) is
\begin{equation}
\label{ecuaciondosdiecinueve}
\rho(\tau,\sigma)=\rho(\tau-\sigma) \ , \quad \varphi(\tau,\sigma)=\varphi(\tau-\sigma) \ .
\end{equation}
These solutions are endowed with the closed-string boundary conditions~(\ref{ecuaciondoscinco}). In addition,~(\ref{ecuaciondoscinco}) supplies the boundary conditions of~$t$ and~$\phi$. The periodicity of~$t$ is trivially satisfied in view of~(\ref{ecuaciondosdiez}). The quasiperiodicity of~$\phi$ imposes an additional constraint on~(\ref{ecuaciondosdiecinueve}). If we use~(\ref{ecuaciondosocho}), the constraint reads
\begin{equation}
\label{ecuaciondosveinte}
2\pi n=-2\int_{0}^{2\pi}\dif\sigma\sinh^2\rho\,\varphi' \ ,
\end{equation}
where we have ignored~$\textnormal{O}(k/J)$ terms.  

The solution~(\ref{ecuaciondosdiecinueve}) encodes the semi-classical limit of those string solutions that are reproducible from the spin chain. Let us describe some of them. The special case~$\rho=0$, where~$\psi$ in the global coordinate system~(\ref{ecuaciondosdos}) and hence~$\varphi=\psi-t$ are undefined, is the solution of the BMN vacuum. The BMN vacuum is the ground state of the spin chain of~\cite{1804.01998,1806.00422}. The solution represents a degenerate world-sheet, which is the time-like geodesic along the centre of~$\adstres$ in the global coordinate system~(\ref{ecuaciondosdos}). 

The solution~(\ref{ecuaciondosdiecinueve}) also comprises pulsating strings~\cite{0001053}, which are classical closed-string solutions. Pulsating strings are written in the chart~(\ref{ecuaciondosdos}) of~$\adstres$;~$\rho$ and~$t$ only depend on~$\tau$, and~$\psi=m\sigma$, where~$m$ is given in~(\ref{ecuaciondoscinco}). The static gauge-fixing condition~(\ref{ecuaciondosdiez}) and the general solution~(\ref{ecuaciondosdiecinueve}) imply~$\rho=\rho_{0}$,~$t=(J/k)\tau$ and~$m=J/k$. Reference~\cite{0001053} proved that geodesics of $\adstres$ written in the chart (\ref{ecuaciondosdos}) map to pulsating strings under spectral flow. The winding number~$m$ of the pulsating strings with respect to~$\psi$ is the integer spectral-flow parameter~$w$, which therefore equals~$J/k$ in the case of our solution. Reference~\cite{0001053} indeed proved that solutions with constant hyperbolic radii arise when~$J/k$ equals~$w$. Let us moreover note that the equality between~$w$ and~$m$ just concerns pulsating strings;~$w$ in general labels spectrally flowed sectors in the world-sheet CFT$_{2}$. We finally stress that these pulsating strings are never degenerate because the winding number~$m=J/k$ is always large.

Solutions with~$\rho=\rho_{0}$ constitute the threshold between `short' and `long' pulsating strings~\cite{0001053}. Short pulsating strings are connected to the principal discrete series of~$\SL$. They arise from time-like geodesics in $\adstres$ (hence the solution of the BMN vacuum) under spectral flow. Long pulsating strings are connected to the principal continuous series of $\SL$. They arise from space-like geodesics in $\adstres$ under spectral flow. Our solutions are compatible with the unitarity bound of~\cite{0001053} on the principal discrete series of~$\SL$, which is discussed in~\cite{1804.01998, 1806.00422}. We prove in Section~\ref{secciontres} that~$J/k$ is bounded from below by~$w$ in the semi-classical limit~$k\rightarrow\infty$; see~(\ref{ecuaciontresquince}) of Section~\ref{secciontres}. Note that solutions with~$\rho=\rho_{0}$ are also compatible with the claim made in~\cite{2010.02782} on the violation of the unitarity bound in the principal continuous series. Compatibility is possible because of the limit~$k\rightarrow\infty$, which permits the saturation of the endpoints of the unitarity bound by exceptional solutions to the Bethe equations.

\section{The \texorpdfstring{$\mathrm{SL}(2,\mathbb{R})$}{SL(2,R)} spin-chain \texorpdfstring{$\sigma$}{sigma}-model from the spin chain}

\label{secciontres}

In this section, we construct the action of the~$\SL$ spin-chain $\sigma$-model starting from the spin chain proposed in~\cite{1804.01998,1806.00422}. We proceed along the lines of~\cite{1905.05533}, where the~$\SU$ spin-chain $\sigma$-model was derived from such a quantum integrable system. To begin, let us briefly review the relevant aspects of the spin chain of~\cite{1804.01998,1806.00422}.

The spin chain of~\cite{1804.01998,1806.00422} encodes the spectrum of the~$\PSU$~WZNW model in a system of Bethe equations. The Bethe equations are built on the transition amplitude of the S-matrix. They determine the set of admissible momenta of the eigenstates of the Hamiltonian called `magnons'. Magnons consist of a linear superposition of creation operators (`oscillators') above the BMN vacuum, which is the ground state of the spin chain. Magnons have definite momentum. Single magnons, in particular, are expressed as a linear superposition each of whose terms involves just one oscillator. The dispersion relation of magnons follows from the imposition of a shortening condition to the Hamiltonian. Such a condition reads
\begin{equation}
\label{ecuaciontresuno}
H^2=\left(\frac{k}{2\pi}P+M\right)^2 \ ,
\end{equation}
where~$H$ denotes the Hamiltonian,~$P$ denotes the momentum operator, and~$M$ is a linear operator that shifts the dispersion relation according to the excitations on which it acts. The Hamiltonian is semi-definite positive owing to the Bogomol'yi-Prasad-Sommerfield bound. Therefore, the dispersion relation involves the positive branch of the absolute value following from~(\ref{ecuaciontresuno}). Magnons are called `chiral' and `anti-chiral' if the expression inside the absolute value of the dispersion relation is positive and negative, respectively. 

Henceforth, we focus on the~$\SL$ sector of the spin chain. `Sector' denotes a choice of the type of oscillator that acts on the BMN vacuum. The type of oscillator is determined by its representation labels under the superisometry algebra of the~$\variedaddos$ background. The various types of oscillators are listed in~\cite{1410.0866} (in the general integrable deformation of~\cite{1209.4049} of the~$\variedaddos$ background with pure NSNS flux).~\footnote{
Reference \cite{1410.0866} listed the canonical creation and annihilation operators in the BMN limit. We denote by `oscillator' the deformation of the creation operator of \cite{1410.0866} along the lines of \cite{0612229} with the same representation labels. The oscillators have not appeared in the bibliography to the best of our knowledge.
}
They have not appeared in the bibliography to the best of our knowledge. The matrix elements of the S-matrix between states with different kinds of oscillator are in general non-trivial. The determination of whether a sector is closed or not is thus elaborate. The semi-classical limit~$k\rightarrow\infty$ allows us to circumvent this problem because it maps sectors to truncations of the classical~$\PSU$ WZNW model. A sector is closed if the corresponding truncation is consistent. It is immediate to check that the~$\SL$ sector is closed when $k\rightarrow\infty$. 

In fact, there are not one but two $\SL$ sectors in the spin chain, $\viz$ the left-handed sector~$\SLL$ and right-handed sector~$\SLR$. Both sectors are present because~$\adstres$ is~$\SL\cong\SLLRdos$ as a permutation coset, and hence the isometry group of the~$\SL$ WZNW model is~$\SLL\times\SLR$. The duplicity is reflected in the eigenvalue~$m$ of~$M$ in~(\ref{ecuaciontresuno}). The BMN vacuum has~$m=0$. Single magnons that transform under~$\SLL$ have~$m=1$; single magnons that transform under~$\SLR$ have~$m=-1$. The dispersion relation of composite magnons follows from these considerations. The important aspect for us is that the restriction of~$m$ reflects the restriction of the spin chain to a sector of given handedness. For the sake of definiteness, we focus on the~$\SLL$ sector of the spin chain.

Moreover, the number of sites~$J$ of the spin chain is bounded~\cite{1806.00422}.~\footnote{
Reference~\cite{1806.00422} identifies~$J$ with the length of the spin chain to introduce the decompactification limit where the S-matrix is definable. The convention is shared by~\cite{0311203}. To clarify LL limit, we identify~$J$ with the number of sites of the spin chain. Our convention is shared by~\cite{1905.05533,0404133,0409086}. Both conventions are related by a rescaling of $\sigma$ in~(\ref{ecuaciontresdieciocho}).} 
In the spin-chain frame in which the spin chain is defined~\cite{1806.00422},~$J$ equals the total angular momentum of the BMN vacuum. The number of sites is constrained by a unitarity bound, which differs among spectrally flowed sectors~\cite{0001053, 1806.00422}. In particular, the unitary bound of the $w$-th spectrally flowed sectors reads
\begin{equation}
\label{ecuaciontresdos}
kw+1\leq J\leq k(w+1)-1 \ ,
\end{equation}
where~$w$ denotes the integer spectral-flow parameter and~$k\in\mathbb{N}$ denotes the level. In writing~(\ref{ecuaciontresdos}), we have further assumed that~$k>1$ and~$w\geq0$. The exclusion of~$k=1$ is unimportant because we concern ourselves with~$k\rightarrow\infty$. In addition, we have made the assumption~$w\geq0$ for simplicity. One may consider~$w<0$ by inverting the sign of~$J$ in~(\ref{ecuaciontresdos}) without major modifications in the remainder of the section. 

In general, the action of a spin-chain $\sigma$-model is built upon coherent states. Coherent states in the spin chain are constructed from the tensor product of~$J$ copies of one-site coherent states. The construction of one-site coherent states is prescribed by the Perelomov procedure (see Appendix~A of~\cite{0404133} for a summary). The procedure requires three elements: a group, a representation thereof, and a reference state. The reference state must be invariant under the action of the Cartan torus of the group up to a phase. We specify these elements as in~\cite{Perelomov:1986tf,0409086}.

The group is~$\SL$. The unitary irreducible representations of~$\SL$ that are non-trivial are infinite-dimensional. These representations are enumerated in, for instance, Subsection~4.1 of~\cite{0001053}. We choose the representation along the lines of~\cite{1905.05533}. Reference~\cite{1905.05533} postulated an ansatz for the representation of one-site coherent states, $\viz$ the~$s=1/2$ fundamental representation of~$\SU$. The equality of the effective action with the result of~\cite{1311.1794} supported the ansatz. Analogously, we postulate that one-site coherent states belong to the~$j=1/2$ lowest-weight principal discrete representation of~$\SL$. The~$j=1/2$ representation of~$\SL$ is realised in each one-site Hilbert space~$\mathcal{H}_{a}$, with~$a=0,...,J-1$. This choice is supported by the final effective action~(\ref{ecuaciontresdiecinueve}), which matches~(\ref{ecuaciondosdiecisiete}) in Section~\ref{secciondos}. We emphasise that the representation of coherent states under~$\SL$ does not coincide with the representation of zeroth-level generators of the current algebra. Instead, we assume the existence of the mapping between our coherent states and states in the world-sheet CFT$_{2}$ of the~$\PSU$ WZNW model. Moreover, we choose the reference state in each~$\mathcal{H}_{a}$ as the state whose isotropy group is maximal. One-site coherent states are thus unambiguously determined~\cite{Perelomov:1986tf,0409086}. They read
\begin{equation}
\label{ecuaciontrestres}
\ket{\vec{n}_{a}}=\sech\rho_{a}\overset{\infty}{\underset{m=0}{\sum}}\e^{-\im m\,\varphi_{a}}\tanh^{m}\rho_{a}\ket{m} \ ,
\end{equation}
where~$\ket{m}$ is an orthonormal basis of~$\mathcal{H}_{a}$. One-site coherent states would present a global phase in general, but~(\ref{ecuaciontrestres}) suffices to recover~(\ref{ecuaciondosdiecisiete}). (Such a global phase was used in~\cite{1905.05533} to match the effective action of~\cite{1311.1794}.) The pair~$\rho_{a}$ and~$\varphi_{a}$ is the discrete counterpart of~$\rho$ and~$\varphi$ in the Lagrangian~(\ref{ecuaciondoscuatro}). The range of both pairs is the same; in fact,~$\rho_{a}$ and~$\varphi_{a}$  will match~$\rho$ and~$\varphi$ under the application of the LL limit. We have defined the short-hand notation~$\vec{n}_{a}$ as
\begin{equation}
\label{ecuaciontrescuatro}
\vec{n}_{a}=[\,\cosh(2\rho_{a}),-\sinh(2\rho_{a})\sin\varphi_{a},\,\sinh(2\rho_{a})\cos\varphi_{a}\,] \ ,
\end{equation}
which labels the one-site coherent state. The vector (\ref{ecuaciontrescuatro}) is assembled from the expectation value of the generators of $\mathfrak{sl}(2,\mathbb{R})$ in the coherent state (\ref{ecuaciontrestres})

A general coherent state in the Hilbert space of the spin chain~$\mathcal{H}=\mathcal{H}_{0}\otimes...\otimes\mathcal{H}_{J-1}$ is
\begin{equation}
\label{ecuaciontrescinco}
\ket{\vec{n}}=\overset{J-1}{\underset{a=0}{\bigotimes}}\ket{\vec{n}_{a}} \ .
\end{equation}
Since the spin chain is closed,~$\vec{n}_{0}$ is identified with~$\vec{n}_{J}$. This identification permits the emergence of closed-string boundary conditions in the continuum limit of the spin chain. In addition, we identify the particular state with every~$\rho_{a}=0$, which consists of~$J$ replicas of~$\ket{0}$, with the BMN vacuum. Coherent states~(\ref{ecuaciontrescinco}) are an overcomplete basis of~$\mathcal{H}$. This fact implies two (correlative) properties which we employ later. The first is the resolution of the identity operator in~$\mathcal{H}$ in the coherent-state basis:
\begin{equation}
\label{ecuaciontresseis}
1=\int\dif\mu[\vec{n}]\ket{\vec{n}}\!\bra{\vec{n}} \ ,
\end{equation}
where~$\dif\mu[\vec{n}]$ is the measure which comprises the product of one-site measures~$\dif\mu[\vec{n}_{a}]$.~\footnote{We write neither~$\dif\mu[\vec{n}]$ nor~$\dif\mu[\vec{n}_{a}]$ explicitly for two reasons. First, the group~$\SL$ is the universal cover of the corresponding Lie group, denoted by~$\SL'$ here. Therefore,~$\varphi$ is not compact, and the usage of the measure of~$\SL'$ employed in~\cite{0409086} is not applicable. (The compact counterpart of~$\varphi$ in~$\SL'$ is introduced in~\cite{0409086}, and~$\varphi$ is decompactified in the final effective action.) Second, the measure of both~$\SL$ and~$\SL'$ is defined at~$j=1/2$ through an analytic continuation of the measure of arbitrary~$j$~\cite{0409086}. These obstructions are not directly relevant for us:~$\dif\mu[\vec{n}_{a}]$ eventually becomes the formal path-integral measure~$[\dif\mu]$ in the path integral~(\ref{ecuaciontresdiecisiete}).} The second property is that coherent states are not orthonormal, but rather satisfy
\begin{equation}
\label{ecuaciontressiete}
\bra{\vec{n}}\!\vec{n}'\rangle=\underset{a=0}{\overset{J-1}{\prod}}\left[\cosh\rho_{a}\cosh\rho'_{a}-\e^{\im\,(\varphi_{a}-\varphi'_{a})}\sinh\rho_{a}\sinh\rho'_{a}\right]^{-1} \ .
\end{equation}

Once we have presented the spin chain and our coherent states, we can start considering the~LL limit of the spin chain. The goal of the~LL limit here is a path-integral representation for a transition amplitude wherein an effective action is identifiable, similar to the approach taken by~\cite{1905.05533}. The starting point is thus the transition amplitude between an initial state~$\ket{\Psi_{1}}$ at~$t=-T/2$ and a final state~$\ket{\Psi_{2}}$ at~$t=T/2$. The parameter~$t$ is the temporal coordinate of the spin chain. The limit~$T\rightarrow\infty$ must be further considered as the time-like direction in the world-sheet is non-compact. Therefore, the transition amplitude is
\begin{equation}
\label{ecuaciontresocho}
Z=\underset{T\rightarrow\infty}{\lim}\bra{\Psi_{2}}\exp(-\im H T)\ket{\Psi_{1}} \ .
\end{equation}
This transition amplitude involves the Hamiltonian~$H$. The action of the Hamiltonian in the spin chain is not directly available. Instead, the action is encoded in the quadratic constraint~(\ref{ecuaciontresuno}). The form of~$H$ is thus not achievable in general; it depends on the specific state in which it acts. To proceed, we follow~\cite{1905.05533}. The initial state~$\ket{\Psi_{1}}$ (or the final state~$\ket{\Psi_{2}}$) is accordingly assumed to be a magnon, $\ie$ an eigenstate of~$H$,~$P$ and~$M$ with eigenvalues~$E$,~$p$ and~$m$, respectively.~\footnote{Reference~\cite{1905.05533} assumes that~$\ket{\Psi_{1}}$ is a coherent state whose parameterisation is implicitly constrained;~\cite{1905.05533} also assumes that~$M$ is diagonal in the~$s=1/2$ representation of~$\SU$. These assumptions resort to the lack of a mapping between coherent states and states in the world-sheet CFT$_{2}$. We take~$\ket{\Psi_{1}}$ as a magnon instead of a coherent state. The magnon is presumably expressible as a linear superposition of coherent states. The point of view is advantageous in that it either clarifies or avoids some of the steps in~\cite{1905.05533}.} (Note that~$E\geq0$ owing to the fact~$H$ is semi-definite positive,~$p$ is quantised since the spin chain is closed, and~$m\geq0$ because~$\ket{\Psi_{1}}$ belongs to the~$\SLL$ sector; these properties are secondary for the LL limit.) Let the dispersion relation of~$\ket{\Psi_{1}}$ be
\begin{equation}
\label{ecuaciontresnueve}
E=-\left(\frac{k}{2\pi}p+m\right) \ ,
\end{equation}
where we have assumed that $\ket{\Psi_{1}}$ is anti-chiral for definiteness. We can replace the Hamiltonian~$H$ by~$E$ in~(\ref{ecuaciontresocho}), and then replace~$E$ with~(\ref{ecuaciontresnueve}). If we lift~$p$ to an operator level,~(\ref{ecuaciontresocho}) reads
\begin{equation}
\label{ecuaciontresdiez}
Z=\underset{T\rightarrow\infty}{\lim}\e^{\im T m}\bra{\Psi_{2}}\exp(\im T(k/2\pi)P)\ket{\Psi_{1}} \ .
\end{equation}
This form is suited to the path-integral representation of the transition amplitude.

To construct a path integral, we have to introduce a partition of~$[-T/2,T/2]$. Let us slice~$[-T/2,T/2]$ in~$N$ subintervals~$[t_{\alpha+1},t_{\alpha}]$ of equal step length~$\Delta t=T/N$. The endpoints of the subintervals are~$t_{\alpha}=(2\alpha-N)T/2N$, with~$\alpha=0,...,N-1$. If we introduce the resolution of the identity~(\ref{ecuaciontresseis}) between the endpoints of every pair of consecutive subintervals,~(\ref{ecuaciontresdiez}) is rephrased as
\begin{equation}
\label{ecuaciontresonce}
Z=\underset{T\rightarrow\infty}{\lim}\int\dif\mu[\vec{n}_{N}]...\int\dif\mu[\vec{n}_{0}]\overline{\Psi_{2}(\vec{n}_{N})}\left(\overset{N-1}{\underset{\alpha=0}{\prod}}\e^{\im \Delta t m}\bra{\vec{n}_{\alpha+1}}\exp(\im \Delta t(k/2\pi)P)\ket{\vec{n}_{\alpha}}\right)\Psi_{1}(\vec{n}_{0}) \ .
\end{equation}
Here,~$\ket{\vec{n}_{\alpha}}=\ket{\vec{n}_{\alpha,0}}\otimes...\otimes\ket{\vec{n}_{\alpha,J-1}}$,~$\Psi_{1}(\vec{n}_{0})=\bra{\vec{n}_{0}}\ket{\Psi_{1}}$ and~$\overline{\Psi_{2}(\vec{n}_{N})}=\bra{\Psi_{2}}\ket{\vec{n}_{N}}$. Note that~$\Psi_{1}(\vec{n}_{0})$ and~$\overline{\Psi_{2}(\vec{n}_{N})}$ are wave functions in the basis of coherent states. The expression~(\ref{ecuaciontresonce}) involves the matrix elements of the anticlockwise shift operator~$U=\exp(\im\epsilon P)$ raised to the power~$(k/2\pi)\Delta t/\epsilon$, where~$\epsilon$ is the spatial step length. In order for~$U$ to be defined on~$\mathcal{H}$, the step length must satisfy
\begin{equation}
\label{ecuaciontresdoce}
\Delta t=\frac{2\pi}{k}\epsilon \ .
\end{equation}
Therefore, the temporal interval of the closed spin chain is discretised. In general, one may write~$\Delta t$ as a positive integer multiple of~(\ref{ecuaciontresdoce}), but~$\Delta t=(2\pi/k)\epsilon$ is obtained when~$[-T/2,T/2]$ is divided into the maximum amount of subintervals. The condition~(\ref{ecuaciontresdoce}) allows us to reformulate~(\ref{ecuaciontresonce}) as
\begin{equation}
\label{ecuaciontrestrece}
\begin{split}
Z=\underset{T\rightarrow\infty}{\lim}\int\dif\mu_{N}...\int\dif\mu_{0}\overline{\Psi_{2}(\vec{n}_{N})}\left(\overset{N-1}{\underset{\alpha=0}{\prod}}\e^{\im \Delta t\, m}\overset{J-1}{\underset{a=0}{\prod}}\left[\cosh\rho_{\alpha+1,a}\cosh\rho_{\alpha,a-1}\vphantom{\e^{\im(\varphi_{\alpha+1,a}-\varphi_{\alpha,a-1})}}\right.\right.\\
\left.\left.-\e^{\im(\varphi_{\alpha+1,a}-\varphi_{\alpha,a-1})}\sinh\rho_{\alpha+1,a}\sinh\rho_{\alpha,a-1}\right]^{-1}\vphantom{\overset{J-1}{\underset{a=0}{\prod}}}\right)\Psi_{1}(\vec{n}_{0}) \ ,
\end{split}
\end{equation}
where we have used
\begin{equation}
\label{ecuaciontrescatorce}
U\ket{\vec{n}_{\alpha}}=\overset{J-1}{\underset{a=0}{\bigotimes}}\ket{\vec{n}_{\alpha,a-1}} \ ,
\end{equation}
and the scalar product~(\ref{ecuaciontressiete}). The expression~(\ref{ecuaciontrestrece}) is amenable to the LL limit. 

For customary spin chains in quantum mechanics, such as the Heisenberg XXX$_{1/2}$ model, the LL limit yields an effective action in the form of a non-linear $\sigma$-model (the reader is referred to~\cite{Fradkin:1991nr} for a standard textbook treatment of the subject). The LL limit is a spatial continuum limit applied to the classical action inside an exact path integral over coherent states. The LL limit is defined by~$\epsilon\rightarrow0$ and~$J\rightarrow\infty$ with the spin-chain length~$R=J\epsilon$ fixed. Under the assumption that coherent states depend analytically on their site labels, the leading contribution within the action in the LL limit yields the non-linear $\sigma$-model. The exact path integral over coherent states follows from a continuum limit. The latter is applied to a transition amplitude with respect to the temporal coordinate~$t$, and precedes the LL limit. This continuum limit is standard in path integrals and consists of~$\Delta t\rightarrow 0$ and~$N\rightarrow\infty$ with~$T=N\Delta t$ fixed under the assumption that coherent states depend analytically on~$t$. In the present case, the application of an analogous approach to~(\ref{ecuaciontrestrece}) is forbidden. The temporal and spatial continuum limits are intertwined in our spin chain. First and foremost, the step lengths~$\Delta t$ and~$\epsilon$ are intertwined as stated by~(\ref{ecuaciontresdoce}). This relationship implies that~$T=2\pi NR/kJ$. If the continuum limits with respect to~$\alpha$ and~$a$ in~(\ref{ecuaciontrestrece}) keep~$T$ and~$R$ respectively finite, the condition~$N/kJ\sim\textnormal{O}(1)$ must be fulfilled. Therefore,~$N\rightarrow\infty$ and~$J\rightarrow\infty$ must be synchronised. Therefore, the LL limit in~(\ref{ecuaciontrestrece}) consists of the simultaneous application of the limit~$\Delta t\rightarrow 0$ and~$N\rightarrow\infty$ with~$T=N\Delta t$ fixed, and~$\epsilon\rightarrow0$ and~$J\rightarrow\infty$ with~$R=J\epsilon$ fixed. Furthermore, the LL limit is a semi-classical limit, $\ie$ the LL limit presupposes~$k\rightarrow\infty$. The inequality~(\ref{ecuaciontresdos}) bounds~$J$ in terms of~$k$ and~$w$. Since the spectral-flow parameter remains finite,~$J\rightarrow\infty$ already implies~$k\rightarrow\infty$ (which in turn implies that~$\Delta t\rightarrow 0$). The counterpart of~(\ref{ecuaciontresdos}) under~$J,k\rightarrow\infty$ is 
\begin{equation}
\label{ecuaciontresquince}
w\le J/k\le w+1 \ .
\end{equation}
The inequality above indicates that the spin chain belongs to the $w$-th spectrally flowed sector in the semi-classical limit ($\cf$ Formula~(59) of~\cite{0001053}). We emphasise that the effective coupling in Section~\ref{secciondos} is the ratio~$k/J$, and hence~(\ref{ecuaciontresquince}) states that the LL limit is accurate in highly spectrally flowed sectors.

In view of the preceding discussion, let us apply the LL limit to~(\ref{ecuaciontrestrece}). If we assume that~$\rho_{\alpha,a}$ and~$\varphi_{\alpha,a}$ depend smoothly on~$\alpha$ and~$a$, and that~$\epsilon$ and~$\Delta t=(2\pi/k)\epsilon$ are small,~(\ref{ecuaciontrestrece}) becomes
\begin{equation}
\begin{split}
\label{ecuaciontresdieciseis}
Z=\underset{T\rightarrow\infty}{\lim}\int\dif\mu_{N}...\int\dif\mu_{0}\overline{\Psi_{2}(\vec{n}_{N})}\overset{N-1}{\underset{\alpha=0}{\prod}}\left(\e^{\im \Delta t\, m}\overset{J-1}{\underset{a=0}{\prod}}\,\left[1+\im\sinh^2\rho_{\alpha,a}(\Delta t\,\dot{\varphi}_{\alpha,a}+\epsilon\varphi'_{\alpha,a})\vphantom{\e^{\im(\varphi_{\alpha+1,a}-\varphi_{\alpha,a-1})}}\right.\right.\\
\left.\left.+\textnormal{O}(\epsilon^2)\vphantom{\e^{\im(\varphi_{\alpha+1,a}-\varphi_{\alpha,a-1})}}\right]\vphantom{\overset{N-1}{\underset{\alpha=0}{\prod}}}\right)\Psi_{1}(\vec{n}_{0}) \ ,
\end{split}
\end{equation}
where~$\dot{\phantom{x}}$ and~$'$ denote derivatives with respect to~$t_{\alpha}$ and~$x_{a}=a\epsilon$, respectively. If we introduce the short-distance cut-off~$\epsilon$ (or, equivalently,~$\Delta t$), we can deem the expression between round brackets the formal product of two continuous products. Accordingly, we can reword~(\ref{ecuaciontresdieciseis}) as the path integral
\begin{equation}
\label{ecuaciontresdiecisiete}
Z=\int [\dif \mu ]\overline{\Psi_{2}(\vec{n}_{\infty})}\e^{\im S}\Psi_{1}(\vec{n}_{-\infty}) \ .
\end{equation}
The path integral involves various elements. First, the classical action. At leading order in~$\epsilon$, the classical action reads
\begin{equation}
\label{ecuaciontresdieciocho}
S=\frac{1}{\epsilon}\int_{-\infty}^{\infty}\dif t\int_{0}^{R}\dif x\left[\frac{\epsilon}{R}m+\sinh^2\rho\left(\dot{\varphi}+\frac{k}{2\pi}\varphi'\right)\right] \ ,
\end{equation}
where~$\rho=\rho(t,x)$ and~$\varphi=\varphi(t,x)$ are the continuous counterparts of~$\rho_{\alpha,a}$ and~$\varphi_{\alpha,a}$. Note we have applied the limit~$T\rightarrow\infty$ with respect to the interval over~$t$. The expression~(\ref{ecuaciontresdiecisiete}) also involves the path-integral measure~$[\dif \mu]$, which is the formal measure that arises from the product of measures~$\dif\mu[\vec{n}_{\alpha}]$. The path integral extends over continuous coherent-state configurations~$\vec{n}=\vec{n}(t,x)$ with respect to~$[\dif \mu]$. As we have previously anticipated, these coherent-state configurations satisfy periodic boundary conditions:~$\vec{n}(t,x+R)=\vec{n}(t,x)$. They are also subject to the asymptotic boundary conditions~$\vec{n}(\pm\infty,x)=\vec{n}_{\pm\infty}(x)$, where~$\vec{n}_{\infty}$ and~$\vec{n}_{-\infty}$ are the continuous counterparts of~$\vec{n}_{N}$ and~$\vec{n}_{0}$, respectively. Finally, the path integral involves~$\Psi_{1}(\vec{n}_{-\infty})$ and~$\overline{\Psi_{2}(\vec{n}_{\infty})}$. These fields are the counterparts of the wave functions~$\Psi_{1}(\vec{n}_{0})$ and~$\overline{\Psi_{2}(\vec{n}_{N})}$ in the LL limit. The energy of the system is determined by~$\Psi_{1}(\vec{n}_{-\infty})$. In a sense, these wave functions act as sources at~$t=\pm\infty$ of the semi-classical solutions to~(\ref{ecuaciontresdiecisiete}). Our point of view here is strongly influenced by the interplay between closed-superstring vertex operator and semi-classical solutions in the~$\correspondencia$ correspondence~\cite{0304139,1002.1716,1005.4516}. However, we shall not pursue this line of thought further.

Finally, we have to introduce a change of variables to retrieve the effective action~(\ref{ecuaciondosdiecisiete}) of Section~\ref{secciondos}. The change of variables is based on the form of the NG action~(\ref{ecuaciondosquince}). To obtain effective action in Section~\ref{secciondos}, we have expanded the NG action in a series with respect to~$k/J$. Apart from an overall factor of~$J$, the $N$-th derivative in both~$\tau$ and~$\sigma$ appears at order~$\textnormal{O}((k/J)^{N})$. We must find the same pattern here. To put~(\ref{ecuaciontresdieciocho}) in the proper form, we set~$\tau=k t$. We also set~$R=1$, $\ie$~$\epsilon=1/J$ and~$\sigma=2\pi x$ and to match the conventions of Section~\ref{secciondos}. In this way, we conclude
\begin{equation}
\label{ecuaciontresdiecinueve}
S=\frac{J}{2\pi}\int_{-\infty}^{\infty}\dif\tau\int_{0}^{2\pi}\dif\sigma\sinh^2\rho\,(\dot{\varphi}+\varphi')\ ,
\end{equation}
which matches the action of the~$\SL$ spin-chain $\sigma$-model~(\ref{ecuaciondosdiecisiete}). 

We must stress that we have omitted the constant contribution of~$m$ in~(\ref{ecuaciontresdieciocho}). Our assumption is that~$m$ (the eigenvalue of the magnon under $M$) is finite in the limit~$k\rightarrow\infty$, and hence its integral is $\textnormal{O}(1/k)$, which is subleading in the large~$J/k$ expansion. In addition, (\ref{ecuaciontresdiecinueve}) is obtained under the restriction to anti-chiral magnons. If we had started from chiral magnons, we would have obtained (\ref{ecuaciontresdiecinueve}) up to the replacement of $\varphi'$ and $-\varphi'$.

A final remark is in order. The spin chain has two~$\SL$ sectors: the~$\SLL$ sector and the~$\SLR$ sector. The relationship of these sectors with coherent states is not direct because coherent states~(\ref{ecuaciontrescinco}) have been postulated rather than mapped from states in the world-sheet CFT$_{2}$. One may wonder how coherent states are connected with the~$\SLL$ sector and the~$\SLR$ sector in the LL limit. The restriction of coherent states to either sector turns out to be reflected in the orientation of~$\varphi$ in~(\ref{ecuaciontresdiecinueve}). To clarify the reasoning behind this statement, let us turn to the chart of~$\adstres$ whereby the canonical quantisation is performed~\cite{1410.0866}. The coordinate system is 
\begin{equation}
\label{ecuaciontresveinte}
\dif s^2=-\left(\frac{z_{1}^2+z_{2}^2+4}{z_{1}^2+z_{2}^2-4}\right)^2\dif t^2+\frac{16}{(z_{1}^2+z_{2}^2-4)^2}(\dif z_{1}^2+\dif z_{2}^2) \ ,
\end{equation}
where $t$ is the global time-like coordinate in (\ref{ecuaciondosdos}). As we have commented in Section~\ref{secciondos},~$t$ is fixed by a gauge-fixing condition prior to the construction of the spin chain. Oscillators in~$\SLL$ give rise to magnons along the direction of~$Z=-z_{2}+\im z_{1}$; oscillators in~$\SLR$ produce magnons along~$\bar{Z}=-z_{2}-\im z_{1}$. The chart~(\ref{ecuaciontresveinte}) is related to~(\ref{ecuaciondosdos}), the global coordinate system of~$\adstres$, as
\begin{equation}
z_{1}=2\tanh \frac{\rho}{2}\sin\psi \ , \quad z_{2}=2\tanh\frac{\rho}{2}\cos\psi \ .
\end{equation} 
Therefore, the clockwise orientation of~$\psi$ corresponds to oscillators transforming under~$\SLL$; the anticlockwise orientation corresponds to oscillators transforming under~$\SLR$. The orientation of~$\psi$ carries over into~$\varphi$ owing to~$\varphi=\psi-t$. Therefore, the restriction of the orientation of~$\varphi$ in~(\ref{ecuaciontresdiecinueve}) suffices to determine coherent states in either sector. In particular,~$\varphi$ advances clockwise in the~$\SLL$ sector that we have considered.

\section{Conclusions}

\label{seccioncuatro}

In this article, we have obtained the effective action of the~$\SL$ WZNW model in the semi-classical limit, which corresponds to a~$\SL$ spin-chain $\sigma$-model. We have computed the effective action from both the classical action of the~$\SL$ WZNW model and the spin chain of \cite{1804.01998,1806.00422}, and proved that the results match. Therefore, the spin chain directly gives rise to the classical~$\SL$ WZNW model in the semi-classical limit. This fact suggests that the representation of the~$\SL$ WZNW model as a spin chain goes beyond the spectral problem analysed in~\cite{1804.01998,1806.00422}. It may then possible to use the spin chain to compute other quantities of the $\SL$ WZNW model. For instance, the spin chain may permit the computation of the correlation functions of~\cite{0111180}, as \cite{1806.00422} already noted. To clarify the scope of coherent states in the spin chain of  \cite{1804.01998,1806.00422} regarding  both the spectral problem and further applications, some question must be answered. We comment on them below. 

The construction of an effective action from the spin chain was based on an ansatz for coherent states. In particular, we have postulated coherent states in the~$j=1/2$ unitary irreducible representation of the principal discrete series of~$\SL$. We have assumed the existence of a mapping between coherent states in the spin chain and states in the world-sheet CFT$_{2}$ of the~$\SL$ WZNW model. In order for the derivation to be complete, an explicit mapping between the two sets of states is needed. The answer may rely on coherent states in the world-sheet CFT$_{2}$. Coherent states were defined in~\cite{0103044} as eigenstates of the lowering operator at the level~$N=1$ of the current algebra. These coherent states consist of an infinite linear superposition of states. Each state belongs to a different negative level of the current algebra. More precisely, the state at level~$-N'$, where~$N'\in\mathbb{N}$, is obtained by applying the rising operator at the level~$N=-1$ to the BMN vacuum~$N'$ times. Coherent states thus defined minimise the Heisenberg uncertainty relation of a pair of `position' and `momentum' operators~\cite{0103044}. (Inequivalent coherent states were constructed in~\cite{0001180}, but they do not minimise this relation in general.) To construct the coherent states defined in Section~\ref{secciontres}, one should endeavour to assemble the coherent states of~\cite{0103044} in states that transform in the aforementioned~$j=1/2$ representation of~$\SL$. An extension of the coherent states of~\cite{0103044} would in fact be needed to embrace all the spectrally flowed sectors of the~$\SL$ WZNW model. Furthermore, the relationship between the sets of coherent states of~\cite{0103044} and Section~\ref{secciontres} would shed light on the completeness of the latter. We emphasise that analogous considerations can be advanced regarding the coherent states of the~$\SU$ WZNW model proposed in~\cite{1905.05533}.

It may be also worth considering subleading corrections to our~$\SL$ spin-chain $\sigma$-model. Even though the lack of a proper characterisation of coherent states a priori forbids computations from the spin chain, it may be still possible to proceed starting from the classical action. For the computation to be meaningful, the~$\SL$ sector must be closed at the order being considered. In the AdS$_{5}$/CFT$_{4}$ correspondence, subleading corrections were obtained in the closed~$\SU$ sector in~\cite{0403120,0406189}. (Reference~\cite{0406189} analysed the~$\SO$ sector instead, to which the $\SU$ sector belongs;~\cite{0405243} proved that the $\SO$ sector is closed if~$J\rightarrow\infty$.) Canonical perturbation theory would supply a systematic framework to address this task~\cite{0406189}. 

Finally, one may also attempt to construct a fermionic spin-chain $\sigma$-model. The simplest fermionic sector is the~$\SUeszett$ sector. In the~$\correspondencia$ correspondence, the~$\SUeszett$ spin-chain $\sigma$-model was constructed in~\cite{0410022,0503185}.~\footnote{Fermionic spin-chain $\sigma$-models embedding the model were also considered in these references and~\cite{0503159,0602007,0704.1460}.} Four~$\SUeszett$ sectors are identifiable in the~$\PSU$ WZNW model~\cite{1410.0866,1806.00422}. (Recall that sector means a choice of the type of oscillator that acts on the BMN vacuum.) New difficulties arise in the derivation from the classical action~\cite{0503185}. For instance, the imposition of an appropriate gauge-fixing condition for the $\kappa$-gauge symmetry of the action, the computation of consistent truncations, and the introduction of field redefinitions to identify `slow' Gra{\ss}mann-odd target-space coordinates. In the spin chain, the procedure may parallel~\cite{1905.05533} and Section~\ref{secciontres}. The steps would thus be the postulation of coherent states in a representation of~$\SUeszett$, and the subsequent derivation of a semi-classical path integral in the LL limit. Being nilpotent, Gra{\ss}mann-odd variables may further clarify the LL limit. Again, a complete derivation would require the explicit connection between coherent states and states in the world-sheet~CFT$_{2}$.

\subsectionfont{\centering}

\subsection*{Acknowledgements}

\noindent The author is grateful to Rafael Hern{\'a}ndez, Juan Miguel Nieto and Natalie A. Yager for comments on the manuscript. This work has been supported through the grant no. PGC2018-095382-B-I00, and by Universidad Complutense de Madrid and Banco Santander through the grant no. GR3/14-A 910770. The author has been supported by Universidad Complutense de Madrid and Banco Santander through the contract no. CT42/18-CT43/18.

\bibliographystyle{bibliographicstyle}

\bibliography{manuscript}{}

\end{document}